\documentclass[fleqn,twoside]{article}
\usepackage[headings]{espcrc2}

\readRCS
$Id: espcrc2.tex,v 1.2 2004/02/24 11:22:11 spepping Exp $
\usepackage{graphicx}
\usepackage[figuresright]{rotating}

\newcommand{\AmS}{{\protect\the\textfont2
  A\kern-.1667em\lower.5ex\hbox{M}\kern-.125emS}}

\usepackage{graphics}
\usepackage{epsfig}

\input myBabarsym
\input definitions

\title{ALEPH Tau Spectral Functions and QCD}

\author{M. Davier\address[LAL]{Laboratoire de l'Acc\'el\'erateur Lin\'eaire, 
        Universit\'e Paris-Sud 11, 91898 Orsay, France},
   A. H\"ocker\address{Physics Division, CERN, 1211 Geneva 23, Switzerland},
        and
        Z. Zhang\addressmark[LAL]}
       
\runtitle{ALEPH Tau Spectral Functions and QCD}
\runauthor{M. Davier, A. H\"ocker, and Z. Zhang}

\begin{document}

\begin{abstract}
Hadronic $\tau$ decays provide a clean laboratory for the precise 
study of quantum chromodynamics (QCD). Observables based 
on the spectral functions of hadronic $\tau$ decays can be related 
to QCD quark-level calculations to determine fundamental 
quantities like the strong coupling constant, quark and gluon condensates.
Using the ALEPH spectral functions and branching ratios, complemented by
some other available measurements, and a revisited analysis of the 
theoretical framework, the value
$\asm = 0.345 \pm 0.004_{\rm exp} \pm 0.009_{\rm th}$ is obtained.
Taken together with the determination of \asZ from the global electroweak
fit, this result leads to the most accurate test of asymptotic freedom:
the value of the logarithmic slope of $\alpha_s^{-1}(s)$ is found to
agree with QCD at a precision of 4\%. The value of \asZ obtained from
$\tau$ decays is 
$\asZ = 0.1215 \pm 0.0004_{\rm exp} \pm 0.0010_{\rm th} \pm 0.0005_{\rm evol}
 =  0.1215 \pm 0.0012~$.

\vspace{1pc}
\end{abstract}

\maketitle

\section{Introduction}

The $\tau$ is the only lepton of the three-generation Standard Model (SM)
that is heavy enough to decay into hadrons. It is therefore 
an ideal laboratory for studying the charged weak hadronic 
currents and QCD. Observables based 
on the spectral functions of hadronic $\tau$ decays can be related 
to QCD quark-level calculations to determine fundamental quantities 
like the strong coupling constant, quark and gluon condensates. We report 
here the results of a QCD analysis of the final ALEPH spectral 
functions~\cite{taubr} using a revisited theoretical framework~\cite{RMP}.

\section{Tau hadronic spectral functions}
\label{sec:tauspecfun}

\subsection{Definitions}
\label{sec:tauspecfun_def}

The \sf\ $v_1$ ($a_1$, $a_0$), where the subscript 
refers to the spin $J$ of the hadronic system, is defined
for a nonstrange ($|\Delta S|=0$) or strange ($|\Delta S|=1$) vector 
(axial-vector) hadronic decay $\tau^- \to {V^-}\nut$ (${A^-}\nut$). 
The \sf\  is obtained from the normalized invariant mass-squared 
distribution $(1/N_{V/A})(d N_{V/A}/d s)$ for a given hadronic mass 
$\sqrt{s}$ multiplied by the appropriate kinematic factor
\beqn
\label{eq:sf}
   v_1(s)/a_1(s) 
   \hspace{-0.2cm}&=&\hspace{-0.2cm}
           \frac{m_\tau^2}{6\,|V_{ud}|^2\,\Sew}\,
              \frac{\BR(\tau^-\to {V^-/A^-}\,\nut)}
                   {\BR(\tau^-\to e^-\,\nueb\nut)} \nonumber \\
  & &      \hspace{-1.5cm}
              \times\,\frac{d N_{V/A}}{N_{V/A}\,ds}\,
              \left[ \left(1-\frac{s}{m_\tau^2}\right)^{\!\!2}\,
                     \left(1+\frac{2s}{m_\tau^2}\right)
              \right]^{-1}\hspace{-0.3cm}, \\[0.2cm]
   a_0(s) 
   \hspace{-0.2cm}&=&\hspace{-0.2cm} 
           \frac{m_\tau^2}{6\,|V_{ud}|^2\,\Sew}\,
              \frac{\BR(\tau^-\to {\pi^-(K^-)}\,\nut)}
                   {\BR(\tau^-\to e^-\,\nueb\nut)} \nonumber \\
  & &      \hspace{-1.5cm} \times\,\frac{d N_{A}}{N_{A}\,ds}\,
              \left(1-\frac{s}{m_\tau^2}\right)^{\!\!-2}\,
              \hspace{-0.3cm},
\label{eq:spect_fun}
\eeqn
where $\Sew$ accounts for electroweak radiative 
corrections~\cite{Marciano:1988}. 
Since CVC is a very good approximation for the nonstrange sector,
the $J=0$ contribution to the nonstrange vector \sf\ is put to zero, 
while the main contributions to $a_0$ are from the pion or kaon poles,
with $(1/N_{A})d N_{A}/ds = \delta (s-m_{\pi,K}^2)$.
They are connected through partial conservation of the axial-vector 
current (PCAC) to the corresponding decay 
constants, $f_{\pi,K}$.
The \sfs\  are normalized by the ratio of the vector/axial-vector 
\bfr\ $\BR(\tau^-\to {V^-/A^-}\nut)$ to the \bfr\ of 
the massless leptonic, \ie, electron, channel. The direct value for 
$\BR_e$ and the two derived values from $\BR_\mu$ and $\tau_\tau$ 
using lepton universality are in good agreement with each other,
providing a consistent and precise combined ALEPH result for the 
electronic branching fraction,
\beq
\label{eq:uni_be} 
  	\BR_e^{\rm uni}=(17.818 \pm 0.032)\%~.
\eeq

Using unitarity and analyticity, the \sfs\ 
are connected to the imaginary part of the two-point  
hadronic vacuum polarization functions
\beqn
\label{eq:correlator}
\lefteqn{ \Pi_{ij,U}^{\mu\nu}=
	\left(-g^{\mu\nu}q^2+q^\mu q^\nu\right)
	\Pi^{(1)}_{ij,U}(q^2) } \nonumber \\
  &  & \hspace{0.9cm} +q^\mu q^\nu\Pi^{(0)}_{ij,U}(q^2)
\eeqn
of vector $(U_{ij}^\mu= V_{ij}^\mu=\qbar_j\gamma^\mu q_i)$
or axial-vector ($U_{ij}^\mu= A_{ij}^\mu=\qbar_j\gamma^\mu\gamma_5 q_i$) 
color-singlet quark currents, and
for time-like momenta-squared $q^2>0$. Lorentz decomposition 
is used to separate the correlation function into its $J=1$ and $J=0$ 
parts. The polarization functions $\Pi_{ij,U}^{\mu\nu}(s)$ have a branch 
cut along the real axis in the complex $s=q^2$ plane. Their imaginary 
parts give the spectral functions defined in~(\ref{eq:sf}), for nonstrange 
quark currents
\beqn
\label{eq:imv}
   \Im\Pi^{(1)}_{u d,V/A}(s)
   &=& \frac{1}{2\pi}v_1/a_1(s)~,  \nonumber \\
   \Im\Pi^{(0)}_{u d,A}(s)
   &=& \frac{1}{2\pi}a_0(s)~.
\eeqn

The analytic vacuum polarization function $\Pi_{ij,U}^{(J)}(q^2)$ 
obeys, up to subtractions, the dispersion relation
\beq
  \label{eq:dispersion}
	   \Pi_{ij,U}^{(J)}(q^2)  =
	     \frac{1}{\pi} \intl_{0}^{\infty}
	     ds\,\frac{{\rm Im}\Pi_{ij,U}^{(J)}(s)}{s-q^2-i\varepsilon}~,
\eeq
where the unknown but in general irrelevant subtraction constants 
can be removed by taking the derivative of $\Pi_{ij,U}(q^2)$. The 
dispersion relation allows one to connect the experimentally accessible
spectral functions to the correlation functions $\Pi_{ij,U}^{(J)}(q^2)$, 
which can be derived from QCD.

\subsection{Inclusive nonstrange spectral functions}
\label{sec:tauspecfun_inclusiveresults}

\subsubsection{Vector and axial-vector spectral functions}

The inclusive $\tau$ vector and axial-vector \sfs\   are shown 
in the upper and lower plots of Fig.~\ref{fig:vasf}, respectively.
The left hand plots give the ALEPH 
results~\cite{aleph_taubr,aleph_vsf,aleph_asf} 
together with its most important exclusive contributions, 
and the right hand plots
compare ALEPH with OPAL~\cite{opal_vasf}. The agreement between the
experiments is satisfying.

The curves in the left hand plots of Fig.~\ref{fig:vasf} represent 
the parton model prediction (dotted) and the 
massless perturbative QCD prediction (solid), assuming the relevant 
physics to be governed by short distances. 
The difference between 
the two curves is due to higher order terms in the strong coupling 
$(\as(s)/\pi)^n$ with $n=1,2,3$. At high energies the 
spectral functions are assumed to be dominated by continuum production,
which locally agrees with perturbative QCD. This asymptotic region
is not yet reached at $s=m_\tau^2$ for the vector and axial-vector
\sfs.

\subsubsection{Inclusive $V \pm A$ spectral functions}

\begin{figure*}
 \centerline{
        \epsfxsize7.8cm\epsffile{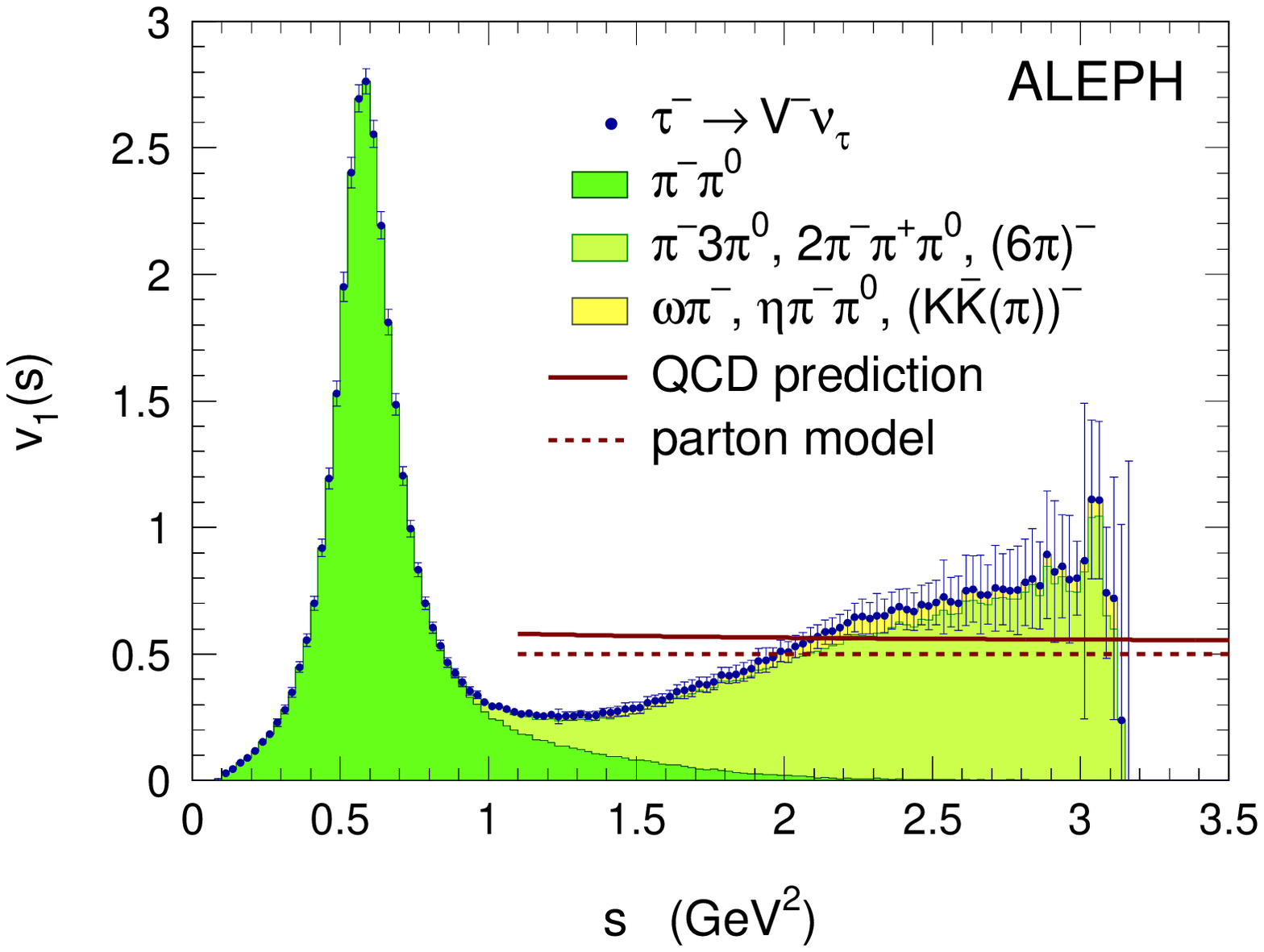}
        \epsfxsize7.8cm\epsffile{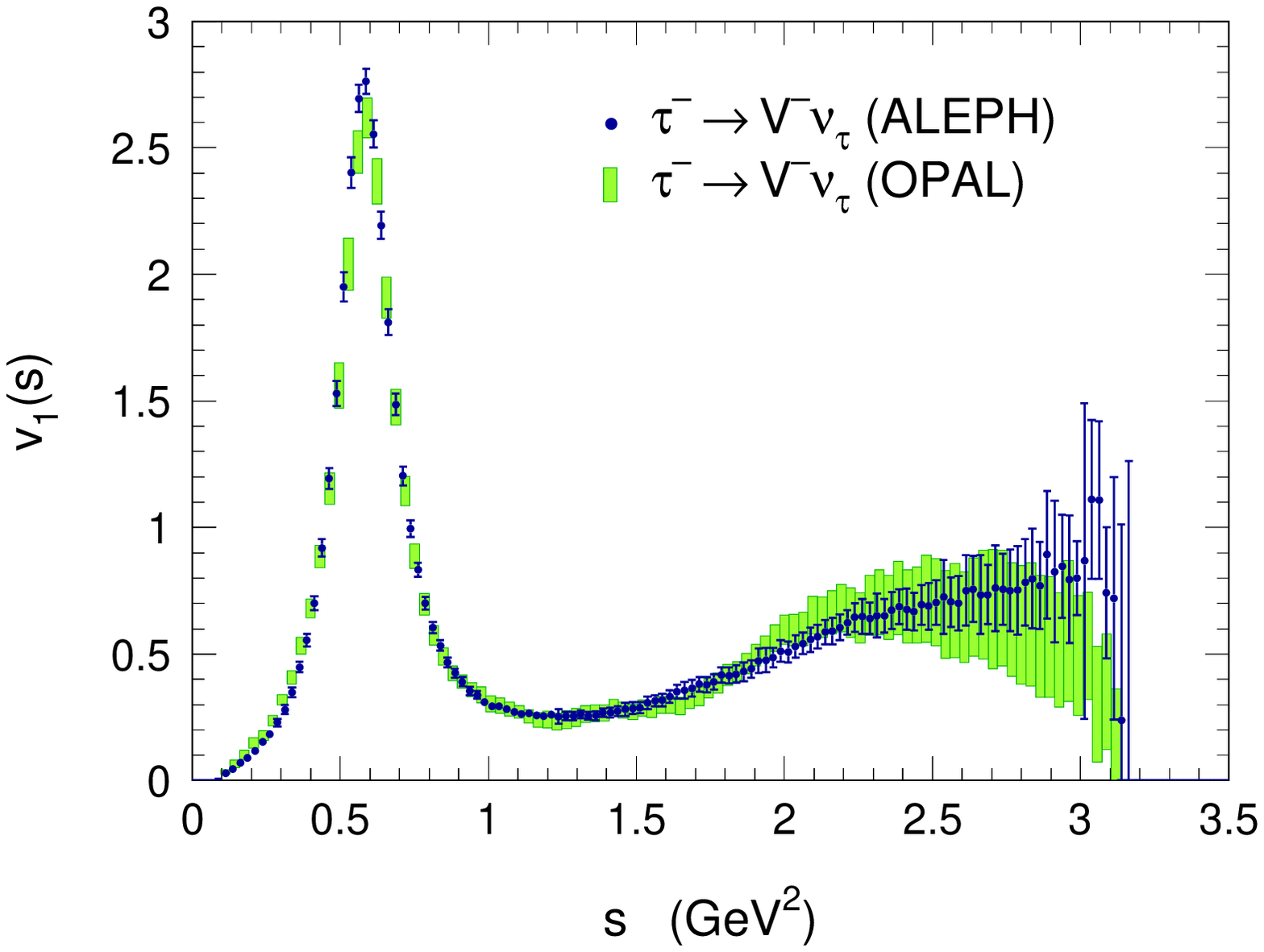}}
  \vspace{0.3cm}
  \centerline{
        \epsfxsize7.8cm\epsffile{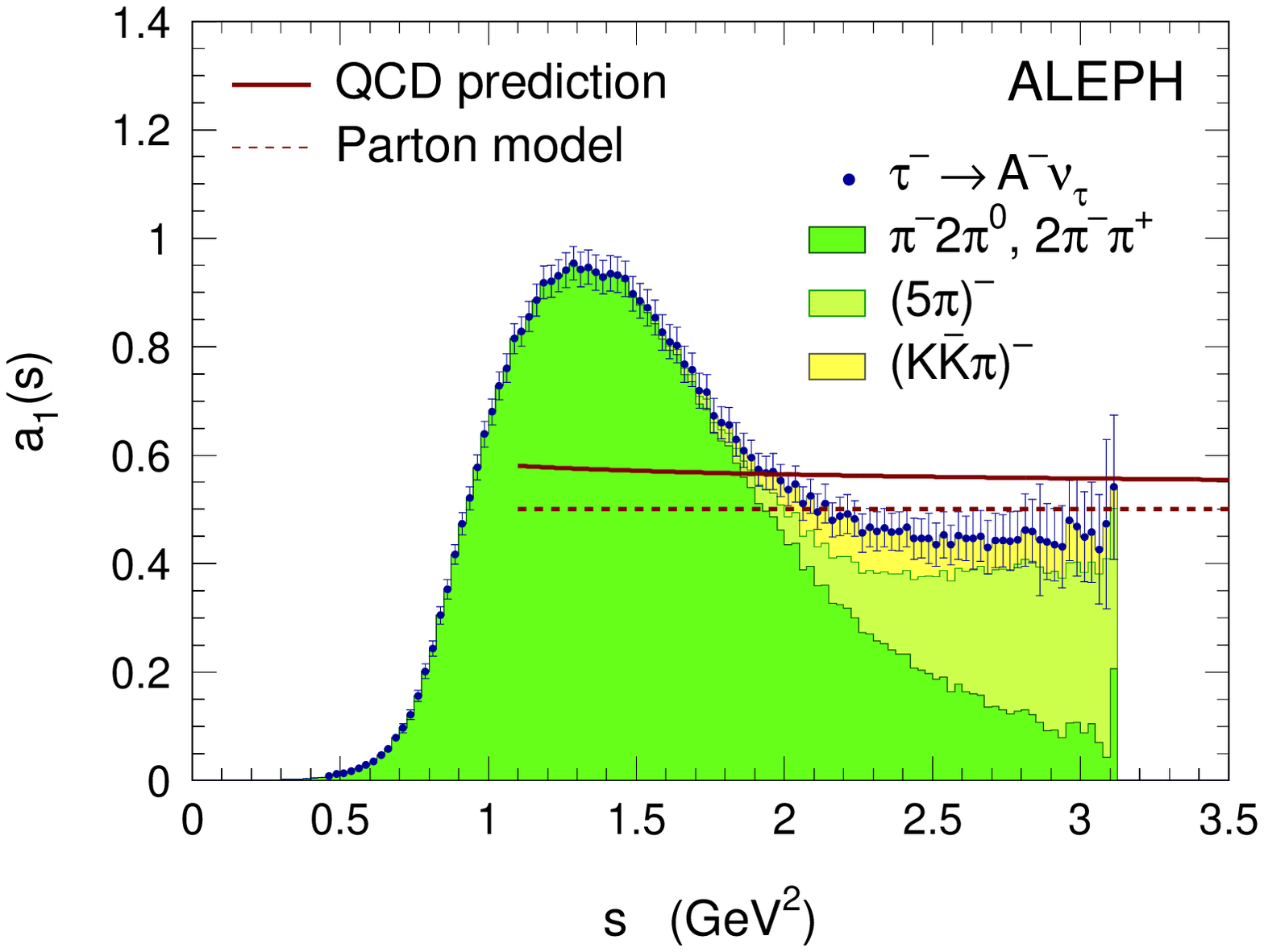}
        \epsfxsize7.8cm\epsffile{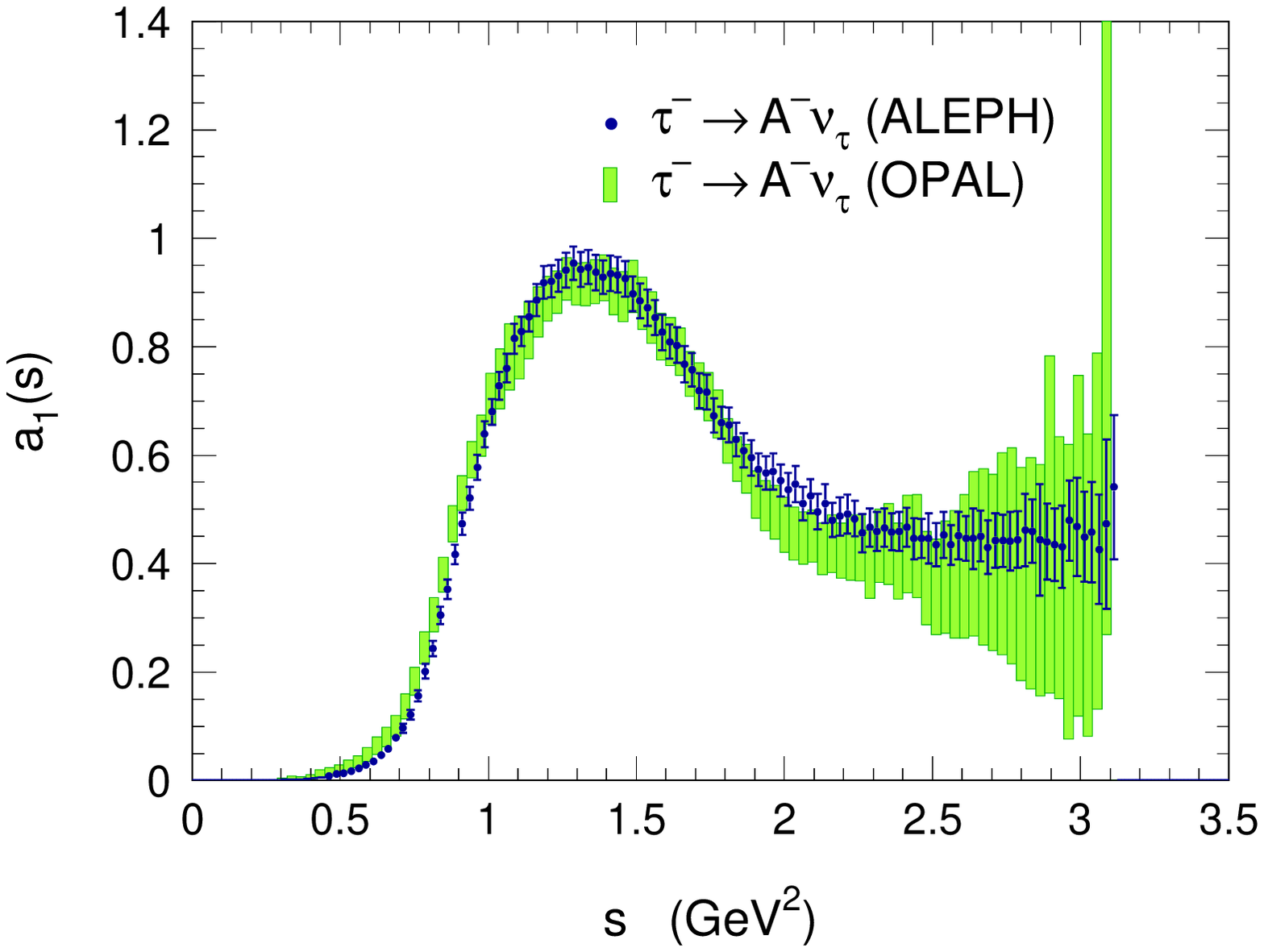}}
  \vspace{-0.8cm}
  \caption[.]{\label{fig:vasf}
        \underline{Left hand plots:} the inclusive vector (upper) and 
	axial-vector (lower) \sfs\   as measured 
	in~\cite{aleph_taubr}. The shaded areas 
	indicate the contributing 
	exclusive $\tau$ decay channels. The curves show the 
	predictions from the parton model (dotted) and from massless 
        perturbative QCD using $\as(M_Z^2)=0.120$ (solid).
	\underline{Right hand plots:} comparison of the inclusive 
	vector (upper) and axial-vector (lower) \sfs\
	obtained by ALEPH and OPAL~\cite{opal_vasf}.}
\end{figure*}

\begin{figure*}
  \centerline{
        \epsfxsize7.8cm\epsffile{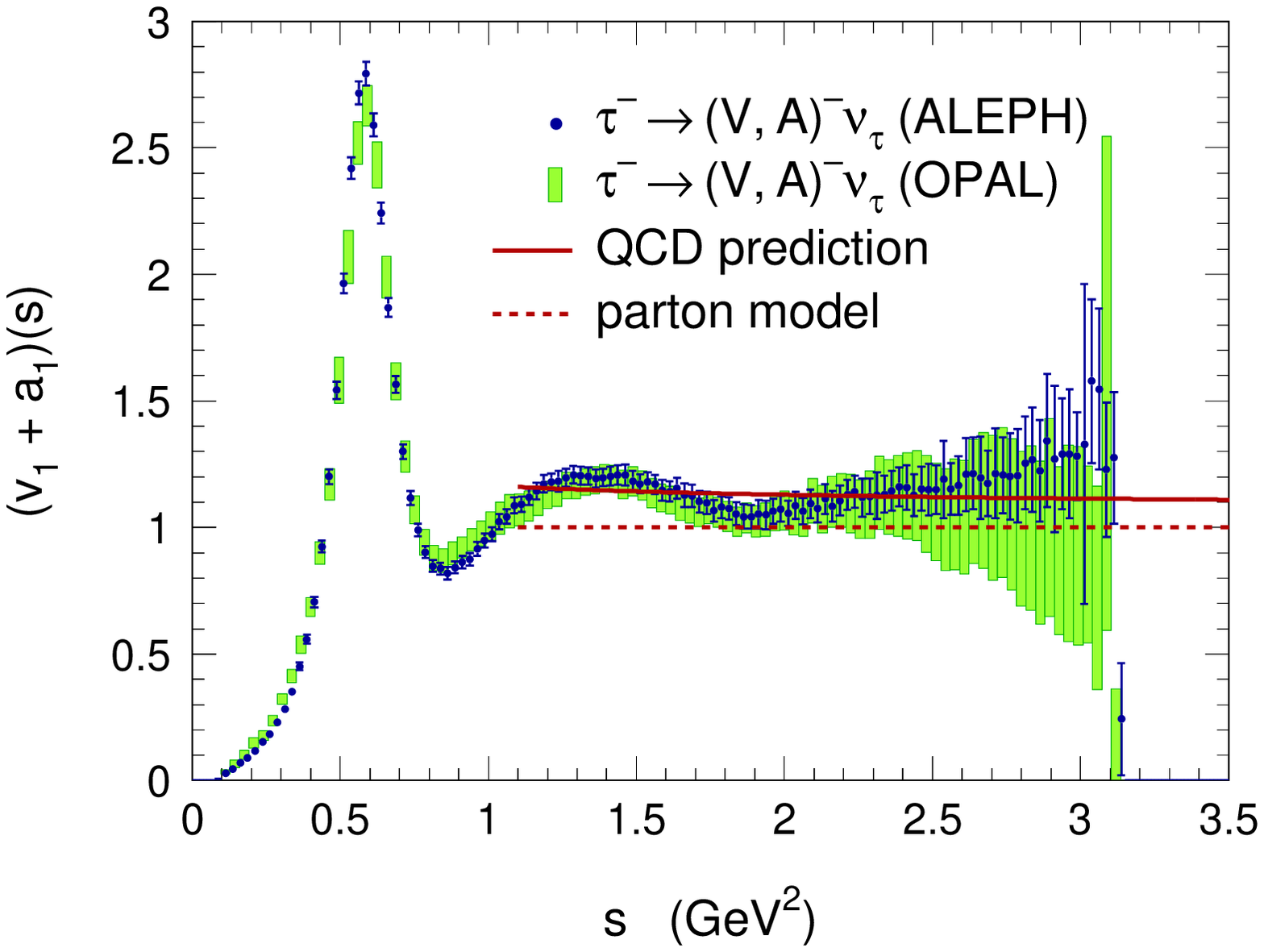}
        \epsfxsize7.8cm\epsffile{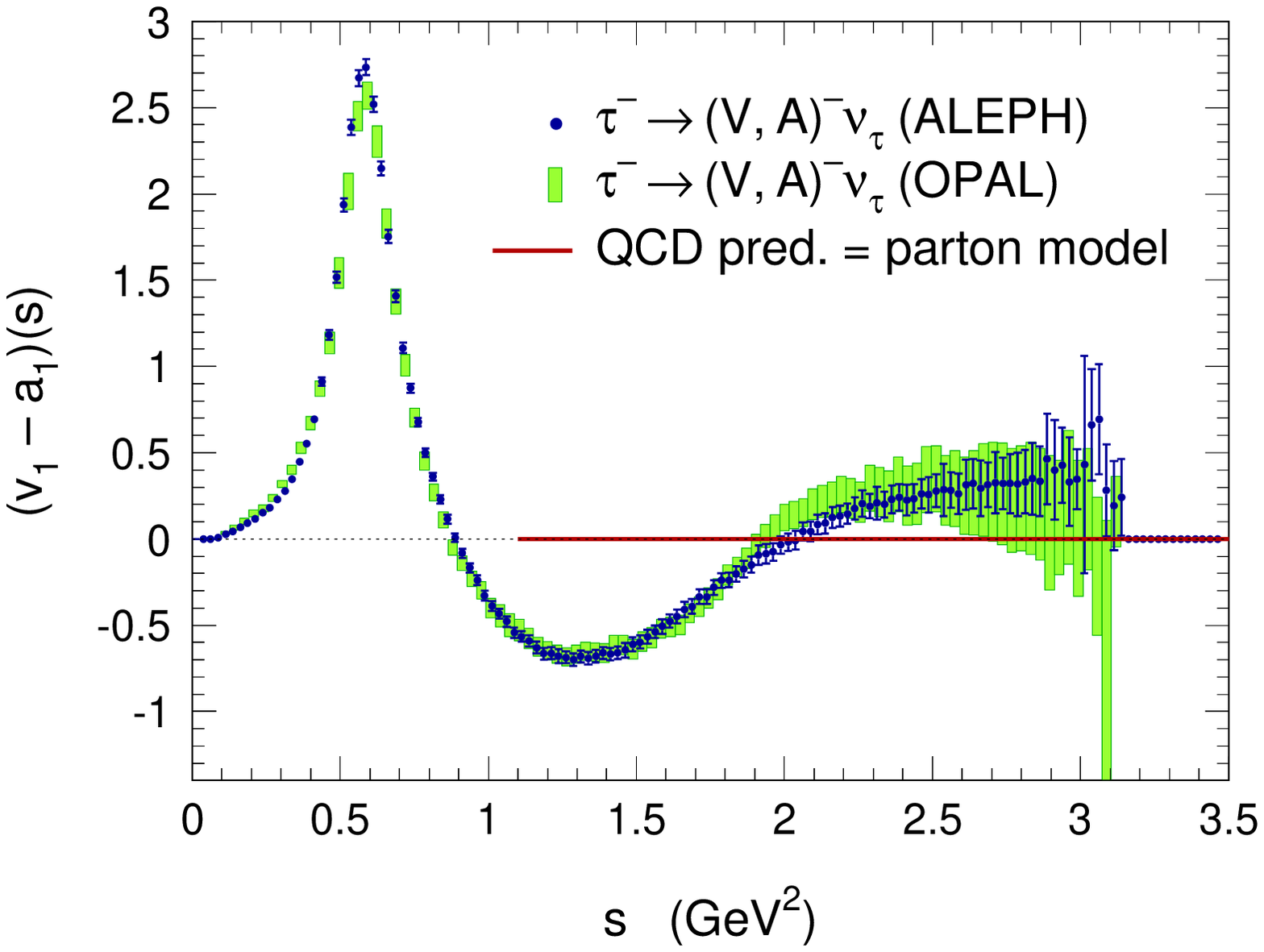}
  }
  \vspace{-0.8cm}
  \caption[.]{\label{fig:vpmasf}
        Inclusive vector plus axial-vector (left) and 
	vector minus axial-vector \sf\  (right) as measured 
	in~\cite{aleph_taubr} (dots with errors bars) 
	and~\cite{opal_vasf} (shaded one standard deviation 
	errors). The lines show the predictions from the 
	parton model (dotted) and from massless 
        perturbative QCD using $\as(M_Z^2)=0.120$ (solid).
	They cancel to all orders in the difference.}
\end{figure*}

For the total $v_1+a_1$ hadronic \sf\  it is not necessary to 
experimentally distinguish whether a given event belongs to
one or the other current.
The one, two and three-pion final states dominate and their exclusive 
measurements are added with proper accounting for anticorrelations
due to the feedthrough. 
The remaining contributing topologies are treated inclusively, \ie, without 
separation of the vector and axial-vector decay modes. This reduces the
statistical uncertainty. The effect of the feed-through between $\tau$
final states on the invariant mass spectrum is described by the Monte 
Carlo simulation and resolution effects are corrected by data unfolding. 
In this procedure the simulated mass distributions are iteratively corrected
using the exclusive vector/axial-vector unfolded mass spectra. 
Also, one does not have to separate the vector/axial-vector currents 
of the $K\Kb\pi$ and $K\Kb\pi\pi$ modes. The $v_1+a_1$ \sfs\  for ALEPH
and OPAL are plotted in the left hand plot of Fig.~\ref{fig:vpmasf}. 
The improvement in precision when comparing to a sum of the two parts 
in Fig.~\ref{fig:vasf} is significant at higher mass-squared values.

One nicely identifies the oscillating behavior of the \sf\  and it is
interesting to observe that, unlike 
the vector/axial-vector \sfs, it does approximately reach the asymptotic 
limit predicted by perturbative QCD at $s\to m_\tau^2$. Also,
the $V+A$ \sf, including the pion pole,
exhibits the features expected from global quark-hadron duality: 
despite the huge oscillations due to the prominent $\pi$, $\rho(770)$, 
$a_1$ and $\rho(1450)$ resonances, the \sf\  qualitatively averages out to 
the quark contribution from perturbative QCD. 

In the case of the $v_1-a_1$ \sf, uncertainties on the $V/A$ 
separation are reinforced due to their relative anticorrelation.
Similarly, the presence of anticorrelations in the branching fractions
between $\tau$ final states with adjacent numbers of pions 
increase the errors. The $v_1-a_1$ \sfs\  for ALEPH and OPAL are 
shown in the right hand plot of Fig.~\ref{fig:vpmasf}. The oscillating 
behavior of the respective $v_1$ and $a_1$ \sfs\  is emphasized and the
asymptotic regime is not reached at $s=m_\tau^2$.
However again, the strong oscillation generated by the hadron resonances
to a large part averages out to zero, as predicted by perturbative QCD.

\section{HADRONIC TAU DECAYS AND QCD}	
\label{sec:qcd}

\subsection{Generalities}

Proposed tests of QCD at the $\tau$ mass 
scale~\cite{narisonpich:1988,braaten:1989,bnp,pichledib}
and the precise measurement of the strong coupling constant \as, 
carried out for the first time by the ALEPH~\cite{aleph_as} and 
CLEO~\cite{cleo_as} collaborations have triggered many theoretical
developments. They concern primarily the perturbative expansion
for which innovative optimization procedures have been suggested. 
Among these are contour-improved (resummed) fixed-order perturbation 
theory~\cite{pert,pivov}, effective charge and minimal 
sensitivity schemes~\cite{grunberg1,pms}, the
large-$\beta_0$ expansion~\cite{beneke,neubert}, and
combinations of these approaches. They mainly distinguish themselves
in how they deal with the fact that the perturbative
series is truncated at an order where the missing part is not
expected to be small. 

One could wonder how $\tau$ decays may at all allow us to learn something about 
perturbative QCD. The hadronic decay of the $\tau$ is dominated by resonant 
single particle final states. The corresponding QCD interactions that bind 
the quarks and gluons into these hadrons necessarily involve long distance 
scales, which are outside the domain of perturbation theory. 
Indeed, it is the inclusive character of 
the sum of all hadronic $\tau$ decays that allows us to probe fundamental
short distance physics. Inclusive observables like
the total hadronic $\tau$ decay rate \Rtau can be accurately 
predicted as function of \asm using perturbative QCD, and including
small nonperturbative contributions within the framework of the Operator 
Product Expansion (OPE)~\cite{svz}. In effect, \Rtau is a doubly inclusive 
observable since it is the result of a summation over all hadronic final 
states at a given invariant mass and further over all masses between 
$m_\pi$ and $m_\tau$. The scale $m_\tau$ lies in a compromise region 
where \asm is large enough so that \Rtau is sensitive to its value, 
yet still small enough so that the perturbative expansion converges 
safely and nonperturbative power corrections are small.

If strong and electroweak radiative corrections are neglected, the 
theoretical parton level prediction for $SU_C(N_C)$, $N_C=3$ reads
\beq
\label{eq:parton}
   R_\tau =
     N_C\left(|V_{ud}|^2 + |V_{us}|^2\right) = 3~,
\eeq
and we can estimate a perturbative correction to this value
of approximately 21\%. One realizes the increase in sensitivity to \as 
compared to the $Z$ hadronic width, where because of the 
three times smaller \asZ the perturbative QCD correction
reaches only about 4\%.

The nonstrange inclusive observable \Rtau can be theoretically separated 
into contributions from specific quark currents, namely vector ($V$) and 
axial-vector ($A$) $\ubar d$ and $\ubar s$ quark currents. It is therefore 
appropriate to decompose
\beq
\label{eq:rtausum}
   \Rtau = \RtauV + \RtauA + \RtauS~,
\eeq
where for the strange hadronic width $\RtauS$ vector and axial-vector
contributions are so far not separated because of the lack of the
corresponding experimental information for the Cabibbo-suppressed modes.
Parton-level and perturbative terms do not distinguish vector and
axial-vector currents (for massless partons). Thus the corresponding 
predictions become 
$R_{\tau,V/A} = (N_C/2)|V_{ud}|^2$ and $\RtauS = N_C|V_{us}|^2$,
which add up to Eq.~(\ref{eq:parton}).

A crucial issue of the QCD analysis at the $\tau$ mass scale is the 
reliability of the theoretical description, \ie, the use of the OPE to 
organize the perturbative and nonperturbative expansions, and the control
of unknown higher-order terms in these series. A reasonable stability test 
is to continuously vary $m_\tau$ to lower values $\sqrt{s_0}\le m_\tau$ for 
both theoretical prediction and measurement, which is possible since the 
shapes of the $\tau$ spectral functions are available. The kinematic 
factor that takes into account the $\tau$ phase space suppression 
at masses near to $m_\tau$ is correspondingly modified so that 
$\sqrt{s_0}$ represents the new mass of the $\tau$. 

\subsection{\boldmath Theoretical prediction of $\Rtau$}\label{sec:rtau_th}

According to Eq.~(\ref{eq:imv}) the absorptive parts of the vector 
and axial-vector two-point correlation functions 
$\Pi^{(J)}_{u d,V/A}(s)$, with the spin $J$ of the hadronic 
system, are proportional to the $\tau$ hadronic \sfs\  with 
corresponding quantum numbers. The nonstrange ratio \RtauVpA
can therefore be written as an integral of these \sfs\  over the 
invariant mass-squared $s$ of the final state hadrons~\cite{bnp}
\beqn
\label{eq:rtauth1}
   \RtauVpA(s_0) \hspace{-0.2cm}&=&\hspace{-0.2cm}
	12\pi \Sew\intl_0^{s_0}
		\frac{ds}{s_0}\left(1-\frac{s}{s_0}
                                    \right)^{\!\!2} \times \\
 & &  \hspace{-2.7cm}  
     \left[\left(1+2\frac{s}{s_0}\right){\rm Im}\Pi^{(1)}(s+i\e)
      \,+\,{\rm Im}\Pi^{(0)}(s+i\e)\right]~, \nonumber
\eeqn
where $\Pi^{(J)}$ can be decomposed as
$\Pi^{(J)}=|V_{ud}|^2\left(\Pi_{ud,V}^{(J)}+\Pi_{ud,A}^{(J)}\right)$.
The lower integration limit is zero because the pion pole is at 
zero mass in the chiral limit.

The correlation function $\Pi^{(J)}$ is analytic in the complex $s$ plane 
everywhere except on the positive real axis where singularities exist.
Hence by Cauchy's theorem, the imaginary part of $\Pi^{(J)}$ is 
proportional to the discontinuity across the positive real axis
\begin{figure}[t]  
  \centerline{\epsfysize6.5cm\epsffile{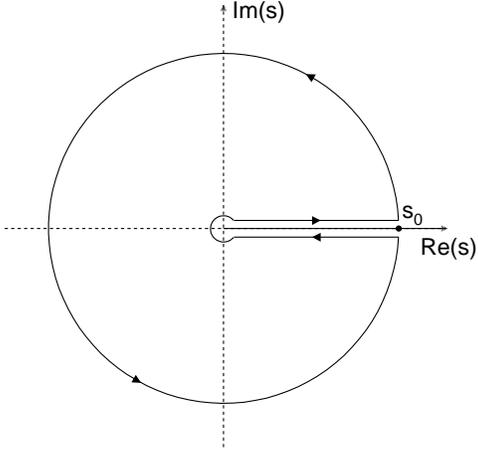}}
  \caption[.]{\label{fig:contour}
              Integration contour for the r.h.s. in Eq.~(\ref{eq:contour}).}
  \vspace{-0.8cm}
\end{figure} 
\beq
\label{eq:contour}
   \intl_0^{s_0}ds\,w(s){\rm Im}\Pi(s) =
   -\frac{1}{2 i}\hm\ointl_{|s|=s_0}\hm\hm\hm ds\,w(s)\Pi(s)~,
\eeq
where $w(s)$ is an arbitrary analytic function, and the contour 
integral runs counter-clockwise around the circle from $s=s_0+i\e$ to 
$s=s_0-i\e$ as indicated in Fig.~\ref{fig:contour}. 

The energy scale $s_0= m_\tau^2$ is large enough that contributions 
from nonperturbative effects are expected to be subdominant and the use 
of the OPE is appropriate. The kinematic factor $(1-s/s_0)^2$ suppresses 
the contribution from the region near the positive real axis where 
$\Pi^{(J)}(s)$ has a branch cut and the OPE validity is restricted 
due to large possible quark-hadron duality violations. 

The theoretical prediction of the vector and axial-vector
ratio \RtauVA can hence be written as
\beqn
\label{eq:delta}
   \RtauVA &=&
     \frac{3}{2}|V_{ud}|^2\Sew\,\, \bigg(1 + \delta^{(0)} + \nonumber \\
 & &   \hspace{-0.5cm} \delta^\prime_{\rm EW} + \delta^{(2,m_q)}_{ud,V/A} + 
     \hm\hm\sum_{D=4,6,\dots}\hm\hm\hm\hm\delta_{ud,V/A}^{(D)}\bigg)~,
\eeqn
with the massless perturbative contribution $\delta^{(0)}$,
the residual non-logarithmic electroweak correction 
$\delta^\prime_{\rm EW}=0.0010$~\cite{braaten}, and the 
dimension $D=2$ {\em perturbative} contribution $\delta^{(2,m_q)}_{ud,V/A}$ 
from quark masses. The term $\delta^{(D)}$ denotes the OPE contributions 
of mass dimension $D$
\beq
\label{eq:ope}
    \delta_{ud,V/A}^{(D)} =
       \hm\hm\hm\sum_{{\rm dim}{\cal O}=D}\hm\hm\hm C_{ud,V/A}(s,\mu)
            \frac{\langle{\cal O}_{ud}(\mu)\rangle_{V/A}}
                 {(-\sqrt{s_0})^{D}}~,
\eeq
where the scale parameter $\mu$ separates the long-distance 
nonperturbative effects, absorbed into the vacuum expectation 
elements $\langle{\cal O}_{ud}(\mu)\rangle$, from the short-distance 
effects that are included in the Wilson coefficients 
$C_{ud,V/A}(s,\mu)$. Note that 
$\delta_{ud,V+A}^{(D)}=(\delta_{ud,V}^{(D)}+\delta_{ud,A}^{(D)})/2$.

\subsubsection{The Perturbative Prediction}\label{sec:pert}

The perturbative prediction used by the experiments follows the
work of~\cite{pert}. Effects from quark masses have been 
calculated in~\cite{pertmass} and are found to be well below
1\% for the light quarks. As a consequence, the contributions from 
vector and axial-vector currents coincide to any given order of 
perturbation theory and the results are flavor independent.

For the evaluation of the perturbative series, it is convenient to 
introduce the analytic Adler function
\beq
\label{eq:adler}
   D(s) \;\equiv\;
      - s\frac{d\Pi(s)}{ds}~.
\eeq
The function $D(s)$, calculated in perturbative QCD within the \MSbar 
renormalization scheme, depends on a non-physical parameter $\mu$ 
occurring as $\ln(\mu^2/s)$. Furthermore it is a function of \as. 
On the other hand, since $D(s)$ is connected to a physical 
quantity, the \sf\  $\Im\Pi(s)$, it cannot depend on the subjective 
choice of $\mu$. This can be achieved if \as\  becomes a function of 
$\mu$ providing independence of $D(s)$ of the choice of $\mu$. 
Nevertheless, in the realistic case of a truncated series, some $\mu$ 
dependence remains and represents an irreducible systematic uncertainty.

The $s$ dependence of the QCD coupling constant is obtained from the
renormalization group equation (RGE)
\beqn
\label{eq:betafun}
	\frac{d a_s}{d \ln s}
	&=& 
	\beta(a_s) = -a_s^2\sum_n \beta_n a_s^n~,
\eeqn
with $a_s=\as/\pi$. Expressed in the \MSbar renormalization scheme 
and for three 
active quark flavors at the $\tau$ mass scale, the $\beta_n$ coefficients 
are known to four loops ($n=3$)~\cite{betafourloop}.

The perturbative expansion of the Adler function can be inferred 
from the 3-loop calculation of the \ee  inclusive cross section ratio 
$R_{\ee}(s)=\sigma(\ee\to{\rm hadrons}\,(\gamma))/\sigma(\ee\to\mu^+\mu^-\,(\gamma))$~\cite{loopbis,loopbisbis}
\beq
\label{eq:adlerpert}
   D(s) =
     \frac{1}{4\pi^2}\sum_{n=0}^\infty \tilde{K}_n(\xi)a_s^{n}(-\xi s)~,
\eeq
where the $\tilde{K}_n(\xi)$ can be expressed~\cite{pert} as function of 
$K_n$ and $\beta_n$ parameters (both known up to order $n=3$), and 
$\xi$, an independent scale parameter. This leads to the perturbative
expansion
\beq 
\label{eq:knan}
   \delta^{(0)} = 
       \sum_{n=1}^5 \tilde{K}_n(\xi) A^{(n)}(a_s)~,
\eeq
with the functions
\beqn
\label{eq:an}
   \hspace{0cm} A^{(n)}(a_s) 
	&=&
      \frac{1}{2\pi i}\hm\ointl_{|s|=s_0}\hm\hm 
      \frac{ds}{s}\,\,\times  \\
  & &   \hspace{-1.5cm}  
       \left[1-2\frac{s}{s_0} + 2\left(\frac{s}{s_0}\right)^{\!\!3}
             - \left(\frac{s}{s_0}\right)^{\!\!4}
       \right]a_s^{n}(-\xi s)~. \nonumber
\eeqn

\subsubsection{Fixed-order perturbation theory (FOPT)}

Inserting the RGE solution for $a_s(s)$ into Eq.~(\ref{eq:knan})  
to evaluate the contour integral, and collecting the terms with 
equal powers in $a_s$ leads to the familiar expression~\cite{pert}
\beq 
\label{eq:kngn}
   \delta^{(0)} = 
       \sum_{n} \left[\tilde{K}_n(\xi) + g_n(\xi)\right]
       \left(\frac{\as(\xi s_0)}{\pi}\right)^{\!\!n}~,
\eeq
where the $g_n$ are functions of $\tilde{K}_{m<n}$ and $\beta_{m<n-1}$, and of
elementary integrals with logarithms of power $m<n$ in the integrand. Setting 
$\xi=1$ and replacing all known $\beta_i$ and $K_i$ coefficients
by their numerical values, Eq.~(\ref{eq:kngn}) simplifies to
\beqn
\label{eq:delta0exp}
   \delta^{(0)}
   &=&
      a_s(s_0)
      + (1.6398 +  3.5625)\,a_s^{2}(s_0) \nonumber \\
   &&
      + (6.371  + 19.995)\,a_s^{3}(s_0) \\
   &&
      +\: (K_4 + 78.003)\,a_s^{4}(s_0) \nonumber\\
   &&
      +\: (K_5 + 14.250\,K_4 - 391.54)\,a_s^{5}(s_0)~, \nonumber
\eeqn
where for the purpose of systematic studies we have kept terms up to 
fifth order. When only two numbers are given in the parentheses, 
the first number corresponds to $K_n$, and the second to $g_n$. 

The FOPT series is truncated at given order 
despite the fact that parts of the higher coefficients $g_{n>4}(\xi)$ are 
known to all orders and could be resummed. These known parts are the 
higher (up to infinite) order logarithmic power terms of the expansion
that are functions of the known $\beta_{n\le3}$ and $K_{n\le3}$ only. 
In effect, beyond the use of the perturbative expansion
of the Adler function~(\ref{eq:adlerpert}), two approximations 
have been used to obtain the FOPT series~(\ref{eq:delta0exp}):
{\em (i)} the RGE~(\ref{eq:betafun}) has been Taylor-expanded 
and terms higher than the given FOPT order have been truncated,
and {\em (ii)} this Taylor expansion is used to predict $a_s(-s)$ on
the entire $|s|=s_0$ contour.

\subsubsection{Contour-improved fixed-order perturbation theory (\FOPTCI)}
\label{sec:cipt}

A more promising approach to the solution of the contour 
integrals~Eq.(\ref{eq:an})
is to perform a direct numerical evaluation by means of single-step
integration and using the solution of the RGE to four loops as 
input for the running $a_s(-\xi s)$ at each integration 
step~\cite{pivov,pert}.
It implicitly provides a partial resummation of the (known) higher 
order logarithmic integrals and improves the convergence of the 
perturbative series. While for instance the third order 
term in the expansion~(\ref{eq:delta0exp}) contributes with $17\%$ 
to the total (truncated) perturbative prediction, the corresponding 
term of the numerical solution amounts to only $6\%$ 
(assuming $\as(m_\tau^2)=0.35$). This numerical solution of 
Eq.~(\ref{eq:knan}) is referred to as {\it contour-improved} 
fixed-order perturbation theory (\FOPTCI) in the following.
Single-step integration also avoids the Taylor approximation of the RGE on 
the entire contour, since $a_s$ is iteratively computed from the previous 
step using the full known RGE.

\subsubsection{\boldmath Other schemes and the value of $K_4$}

Other approaches for evaluating the perturbative prediction have been
presented, such as the effective charge perturbation theory (ECPT, 
see for instance Refs.~\cite{grunberg1,maxwellrs1}) and the large-$\beta_0$
expansion~\cite{beneke,neubert}. Whereas these methods are of considerable 
theoretical interest, they are not suited for precision analyses~\cite{RMP}.

ECPT has been largely used to estimate the value of the first uncalculated 
term $K_4$~\cite{k4_pms}. Another approach to estimating $K_4$~\cite{k4_fld}
is, within CIPT, to enhance the sensitivity to higher order 
perturbative terms by reducing the renormalization scale $\xi$. 
Both methods yield $K_4 \sim 27$, but it has been shown~\cite{RMP}
that the precision of this estimate is seriously limited ($\sim$ 100 \%) 
by the lack of knowledge of the unknown higher order parameters in the
perturbative series ($K_5, \dots$).

Significant efforts are underway
with the goal to calculate the $K_4$ coefficient. Although the large number 
of five-loop diagrams that are needed to calculate the two-point current
correlator at this order may appear discouraging, the results 
on two gauge invariant subsets are already available. The subset of order 
${\cal O}(\alpha_s^4n_f^3)$ was evaluated long ago through the summation of 
renormalon chains~\cite{beneke1993}, while the much harder subset 
${\cal O}(\alpha_s^4n_f^2)$ was recently calculated~\cite{baikov_a4}.
Following these investigations, the value $K_4=25\pm25$ is used in our
analysis.

\subsubsection{Comparison of the perturbative methods}
\label{sec:pert_comp}

To study the convergence of the 
perturbative series, we give in Table~\ref{tab:intsol} the contributions 
of the different orders in PT to $\delta^{(0)}$ for the various approaches
using $\as(m_\tau^2)=0.35$. A geometric growth, $K_n \sim K_{n-1}^2/K_{n-2}$,
is assumed for all unknown PT and RGE coefficients. In the case of CIPT 
the results are given for the various techniques used to evolve $\as(s)$.

\begin{table*}[t]
  \caption[.]{\label{tab:intsol}
              Massless perturbative contribution to $\Rtau(m_\tau^2)$ 
	      for the various methods considered, and at orders 
              $n\ge1$ with $\asm=0.35$. The value of $K_4$ is 
              set to 25, while all unknown higher order $K_{n>4}$ and 
	      $\beta_{n>3}$ coefficients are assumed to follow a geometric 
	      growth. Details are given in Ref.~\cite{RMP}.}
\begin{center}
\setlength{\tabcolsep}{0.0pc}
\begin{tabular*}{\textwidth}{@{\extracolsep{\fill}}lrrrrrrrr} 
\hline\noalign{\smallskip}
  	& \mc{7}{c}{$\delta^0$} \\
Pert. Method	
        & $n=1$	 & $n=2$  & $n=3$   & $(n=4)$ & $(n=5)$ & $(n=6)$   & $\sum_{n=1}^4$ & $\sum_{n=1}^6$ \\
\noalign{\smallskip}\hline\noalign{\smallskip}
FOPT ($\xi=1$)
	& 0.1114 & 0.0646 & 0.0365  & 0.0159  & 0.0010  & $-0.0086$ & 0.2283      & 0.2208 \\
\FOPTCI (Taylor RGE, $\xi=1$)         	
	& 0.1573 & 0.0317 & 0.0126  & 0.0042  & 0.0011  & 0.0001    & 0.2058      & 0.2070 \\
\FOPTCI (full RGE, $\xi=1$)           	
	& 0.1524 & 0.0311 & 0.0129  & 0.0046  & 0.0013  & 0.0002    & 0.2009      & 0.2025 \\
\FOPTCI (full RGE, $\xi=0.4$)           	
	& 0.2166 &$-0.0133$&0.0006  &$-0.0007$& 0.0010  &$-0.0007$  & 0.2032      & 0.2048 \\
ECPT
	& 0.1442 & 0.2187 &$-0.1195$&$-0.0344$&$-0.0160$&$-0.0120$  & 0.2090      & 0.1810 \\
Large-$\beta_0$ expansion
	& 0.1114 & 0.0635 & 0.0398  & 0.0241  & 0.0155  & 0.0093    & 0.2388      & 0.2636 \\
\noalign{\smallskip}\hline
\end{tabular*}
  \end{center}
\end{table*}
Faster convergence is observed for CIPT
compared to FOPT yielding a significantly smaller error associated with 
the renormalization scale ambiguity. Our coarse extrapolation
of the higher order coefficients could indicate that minimal sensitivity 
is reached at $n\sim5$ for FOPT, while the series further converges
for CIPT. Although the Taylor expansion in the CIPT integral exhibits 
significant deviations from the exact solution on the integration circle, 
the actual numerical effect from this 
on $\delta^{(0)}$ is small (\cf\  second and third column in 
Table~\ref{tab:intsol}). The 
convergence of the ECPT series is much worse than for FOPT and CIPT.
Consequently, the difference between truncation at $n=4$ and $n=6$ may 
be significant. A similar instability may occur for the 
large-$\beta_0$ expansion.

The CIPT series is found to be better behaved than FOPT (as well as ECPT)
and is therefore to be preferred for the numerical analysis of the $\tau$
hadronic width. As a matter
of fact, the difference in the result observed when using a Taylor 
expansion and when truncating the perturbative series after integrating
along the contour (FOPT) with the exact result at given order (CIPT) 
exhibits the incompleteness of the perturbative series. However, 
it is even worse than that since large known coefficients are neglected
in FOPT so that the difference between CIPT and FOPT may actually overstate
the perturbative truncation uncertainty (certainly it is not a good measure
of the latter uncertainty). This can be verified by studying 
the behavior of this difference for the various orders in perturbation 
theory given in Table~\ref{tab:intsol}. The CIPT-vs.-FOPT discrepancy 
increases with the addition of each order, up to order four where a 
maximum is reached. Adding the fifth order does not reduce the effect,
and only beyond fifth order the two evaluations may become asymptotic
to each other.
As a consequence varying the unknown higher order coefficients {\em and} using 
the difference between FOPT and CIPT as indicator of the theoretical 
uncertainties overemphasizes the truncation effect. 

\subsection{Results} \label{sec:moment}
It was shown in~\cite{pichledib} that one can exploit the shape of the 
\sfs\  to obtain additional constraints on \assz and---more 
importantly---on the nonperturbative effective operators. The 
{\em $\tau$ spectral moments} at $s_0=m_\tau^2$ are defined by
\beq
\label{eq:moments}
   R_{\tau,V/A}^{k\l} =
       \intl_0^{m_\tau^2} ds\,\left(1-\frac{s}{m_\tau^2}\right)^{\!\!k}\!
                              \left(\frac{s}{m_\tau^2}\right)^{\!\!\l}
       \frac{dR_{\tau,V/A}}{ds}
\eeq
where $R_{\tau,V/A}^{00}=R_{\tau,V/A}$. The factor $(1-s/m_\tau^2)^k$ 
suppresses the integrand at the crossing of the positive real axis where the 
validity of the OPE less certain and the experimental accuracy 
is statistically limited. Its counterpart $(s/m_\tau^2)^\l$ projects upon
higher energies. The spectral information is used to fit simultaneously 
\asm  and the effective operators $\GG$, 
$\rho\as\langle\qbar q\rangle^2$ and $\langle{\cal O}_D\rangle$ for dimension
$D=4$, $6$ and $8$, respectively. Due to the large correlations between the
differently weighted spectral integrals,
only five moments are used as input to the fit.

In analogy to \Rtau, the contributions to the moments originating from 
perturbative and nonperturbative QCD are decomposed through the OPE.

\subsubsection{\boldmath The ALEPH determination of $\as(m_\tau^2)$ and 
               nonperturbative contributions}
\label{sec:qcd_as_fit}

Combined fits to experimental spectral moments and the extraction of
$\as(m_\tau^2)$ together with the leading nonperturbative operators have 
been performed by ALEPH, CLEO and 
OPAL~\cite{opal_vasf,aleph_taubr,aleph_asf,aleph_as,cleo_as} using 
similar strategies and inputs.
This analysis uses the final and complete data on branching fractions and 
spectral functions from ALEPH~\cite{aleph_taubr}, yielding
\beqn
    R_{\tau,V+A}   &=& 3.482 \,\pm\, 0.014~, \\
    R_{\tau,V}     &=& 1.787 \,\pm\, 0.011\, \pm\, 0.007~, \\
    R_{\tau,A}     &=& 1.695 \,\pm\, 0.011\, \pm\, 0.007~, \\
    R_{\tau,V-A}   &=& 0.092 \,\pm\, 0.018\, \pm\, 0.014~,
\eeqn
where the second error originates from the $V/A$ separation in final states
with a $K\overline{K}$ pair, fully anticorrelated beetwen $R_{\tau,V}$ and 
$R_{\tau,A}$.
To reduce the model dependence of the analysis, one fits
simultaneously the nonperturbative operators, which 
is possible since the correlations between these and $\as$ turn out to be 
small enough. The main theoretical uncertainties are due to
$K_4$ (25$\pm$25) and to the renormalization scale, 
which is varied around $m_\tau$ from 1.1 to $2.5\gev$
(the variation over half of the range taken as systematic uncertainty).

\begin{table*}[t]
  \caption[.]{\label{tab_asresults}
	Results~\cite{aleph_taubr} for \asm and the nonperturbative 
	contributions for vector, axial-vector and $V+A$ combined
	fits using the corresponding experimental spectral moments 
	as input parameters. Where two errors are given the first is
	experimental and the second theoretical. The $\delta^{(2)}$ term 
	is theoretical only with quark masses varying within their 
	allowed ranges (see Ref.~\cite{RMP}). 
        The quark condensates in the $\delta^{(4)}$ 
	term are obtained from PCAC, while the gluon condensate is 
	determined by the fit. The total nonperturbative contribution is 
	the sum $\delta_{\rm NP}=\delta^{(4)}+\delta^{(6)}+\delta^{(8)}$.
	Full results are listed only for the \FOPTCI\ perturbative 
        prescription, except for \asm\ where the results using both \FOPTCI\ 
        and FOPT are given (See Ref.~\cite{taubr} for the complete results).}
  \begin{center}
{\small
\setlength{\tabcolsep}{0.0pc}
\begin{tabular*}{\textwidth}{@{\extracolsep{\fill}}lccc} 
\hline\noalign{\smallskip}
  Parameter     &Vector ($V$) &Axial-Vector ($A$)&  $V\,+\,A$\\
\noalign{\smallskip}\hline\noalign{\smallskip}
 \asm  (\FOPTCI) &  $0.355\pm0.008\pm0.009$  
                 &  $0.333\pm0.009\pm0.009$   
                 &  $0.350\pm0.005\pm0.009$   
\\
 \asm  (FOPT)    &  $0.331\pm0.006\pm0.012$  
                 &  $0.327\pm0.007\pm0.012$   
                 &  $0.331\pm0.004\pm0.012$ \\
\noalign{\smallskip}\hline\noalign{\smallskip}
  $\delta^{(2)}$ (\FOPTCI)    & $(-3.3\pm3.0) \times 10^{-4}$
                              & $(-5.1\pm3.0) \times 10^{-4}$
                              & $(-4.4\pm2.0) \times 10^{-4}$
 \\
\noalign{\smallskip}\hline\noalign{\smallskip}
 $\GG$ ($\gev^4$) (\FOPTCI)   &  $(0.4\pm0.3) \times 10^{-2}$  
                                &  $(-1.3\pm0.4) \times 10^{-2}$   
                                &  $(-0.5\pm0.3) \times 10^{-2}$   
\\ 
\noalign{\smallskip}\hline\noalign{\smallskip}
  $\delta^{(4)}$ (\FOPTCI)    & $(4.1\pm1.2) \times 10^{-4}$
                              & $(-5.7\pm0.1) \times 10^{-3}$
                              & $(-2.7\pm0.1) \times 10^{-3}$
 \\      
\noalign{\smallskip}\hline\noalign{\smallskip}
 $\delta^{(6)}$ (\FOPTCI) &  $(2.85\pm0.22) \times 10^{-2}$  
                          &  $(-3.23\pm0.26) \times 10^{-2}$   
                          &  $(-2.1\pm2.2) \times 10^{-3}$   
\\
\noalign{\smallskip}\hline\noalign{\smallskip}
 $\delta^{(8)}$ (\FOPTCI) &  $(-9.0\pm0.5) \times 10^{-3}$  
                          &  $(8.9\pm0.6) \times 10^{-3}$   
                          &  $(-0.3\pm4.8) \times 10^{-4}$   
\\
\noalign{\smallskip}\hline\noalign{\smallskip}
  Total $\delta_{\rm NP}$ (\FOPTCI)    & $(1.99\pm0.27) \times 10^{-2}$  
                                       & $(-2.91\pm0.20) \times 10^{-2}$
                                       & $(-4.8\pm1.7) \times 10^{-3}$
 \\
\noalign{\smallskip}\hline\noalign{\smallskip}
 $\chi^2/$DF (\FOPTCI)       & 0.52            & 4.97            & 3.66    
 \\ 
\noalign{\smallskip}\hline
  \end{tabular*}
}
  \end{center}
\end{table*}

The fit results are given in Table~\ref{tab_asresults}.
There is a remarkable agreement within statistical errors between the 
\asm  determinations using the vector and axial-vector data.
This provides an important consistency check of the results, since
the two corresponding spectral functions are experimentally independent and
manifest a quite different resonant behavior. However it must be mentioned
that the $\as(m_\tau^2)$ determination using either the $V$ and $A$ spectral 
functions is more dependent on the validity of the OPE approach since their 
nonperturbative contributions are significantlty larger than for $V+A$.
Indeed the leading nonperturbative contributions of dimension $D=6$ and 
$D=8$ approximately cancel in the inclusive sum. This cancellation of the 
nonperturbative terms increases the confidence in the \asm determination 
from the inclusive $V+A$ observables. Averaging CIPT and FOPT, the
result quoted by ALEPH is
\beq
  \as(m_\tau^2) = 0.340 \pm 0.005_{\rm exp} \pm 0.014_{\rm th}~,
\eeq
The gluon condensate is 
determined by the first $k=1$, $\l=0,1$ moments, which receive 
lowest order contributions. The values obtained in 
the $V$ and $A$ fits are not very consistent, which could indicate 
problems in the validity of the OPE approach used once the 
nonperturbative terms become significant. Taking the value obtained in 
the $V+A$ fit, where nonperturbative effects are small, 
and adding as systematic uncertainties half of the difference between 
the vector and axial-vector fits as well as between 
the \FOPTCI\ and FOPT results, ALEPH measures the gluon condensate to be
\beq
\label{eq:gluon_cond}
	\GG=(0.001\pm0.012)\gev^4.
\eeq
This result does not provide evidence for a nonzero gluon condensate, but it 
is consistent with and has comparable accuracy to the independent value 
obtained using charmonium sum rules and \ee data in the charm 
region, $(0.011\pm0.009)\gev^4$ in a combined determination with 
the $c$ quark mass~\cite{ioffe}. 

The approximate cancellation of the nonperturbative contributions
in the $V+A$ case was 
predicted~\cite{bnp} for $D=6$ assuming vacuum
saturation for the matrix elements of four-quark operators, which yields
$\delta^{(6)}_V /\delta^{(6)}_A =-7/11=-0.64$, in fair agreement with
the result $-0.90\pm0.18$. The estimate~\cite{bnp} for 
$\delta^{(6)}_V = (2.5 \pm 1.3) \times 10^{-2}$ agrees with 
the experimental result. 

The total nonperturbative $V+A$ correction, 
$\delta_{{\rm NP},V+A}=(-4.3\pm1.9) \times 10^{-3}$, is an
order of magnitude smaller than the corresponding values in the $V$ and 
$A$ components, $\delta_{{\rm NP},V}=(2.0\pm0.3) \times 10^{-2}$ and 
$\delta_{{\rm NP},A}=(-2.8\pm0.3) \times 10^{-2}$.

\subsubsection{\boldmath Running of $\as(s)$ below $m_\tau^2$}
\label{sec:qcd_as_running}
Using the \sfs, one can simulate the physics of a hypothetical 
$\tau$ lepton with a mass $\sqrt{s_0}$ smaller than $m_\tau$
through Eq.~(\ref{eq:rtauth1}). Assuming quark-hadron duality, 
the evolution of $R_\tau(s_0)$ provides a direct test of the running 
of $\as(s_0)$, governed by the RGE $\beta$-function. On the other 
hand, it is also a test of the stability of the OPE approach 
at small scales. The studies performed in this section employ only 
\FOPTCI. 

\begin{figure*}[t]
   \centerline{
        \epsfxsize8.3cm\epsffile{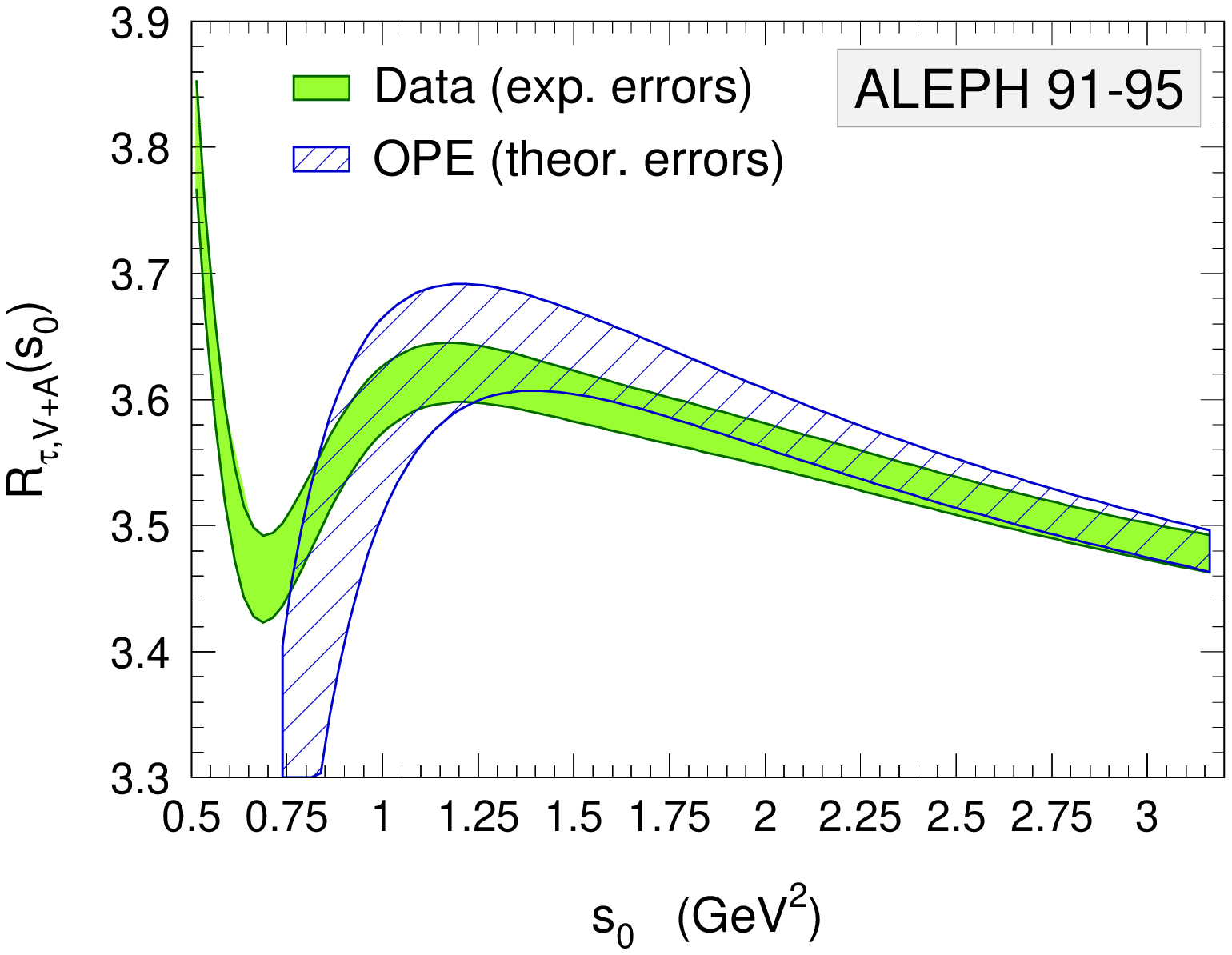}
        \epsfxsize8.3cm\epsffile{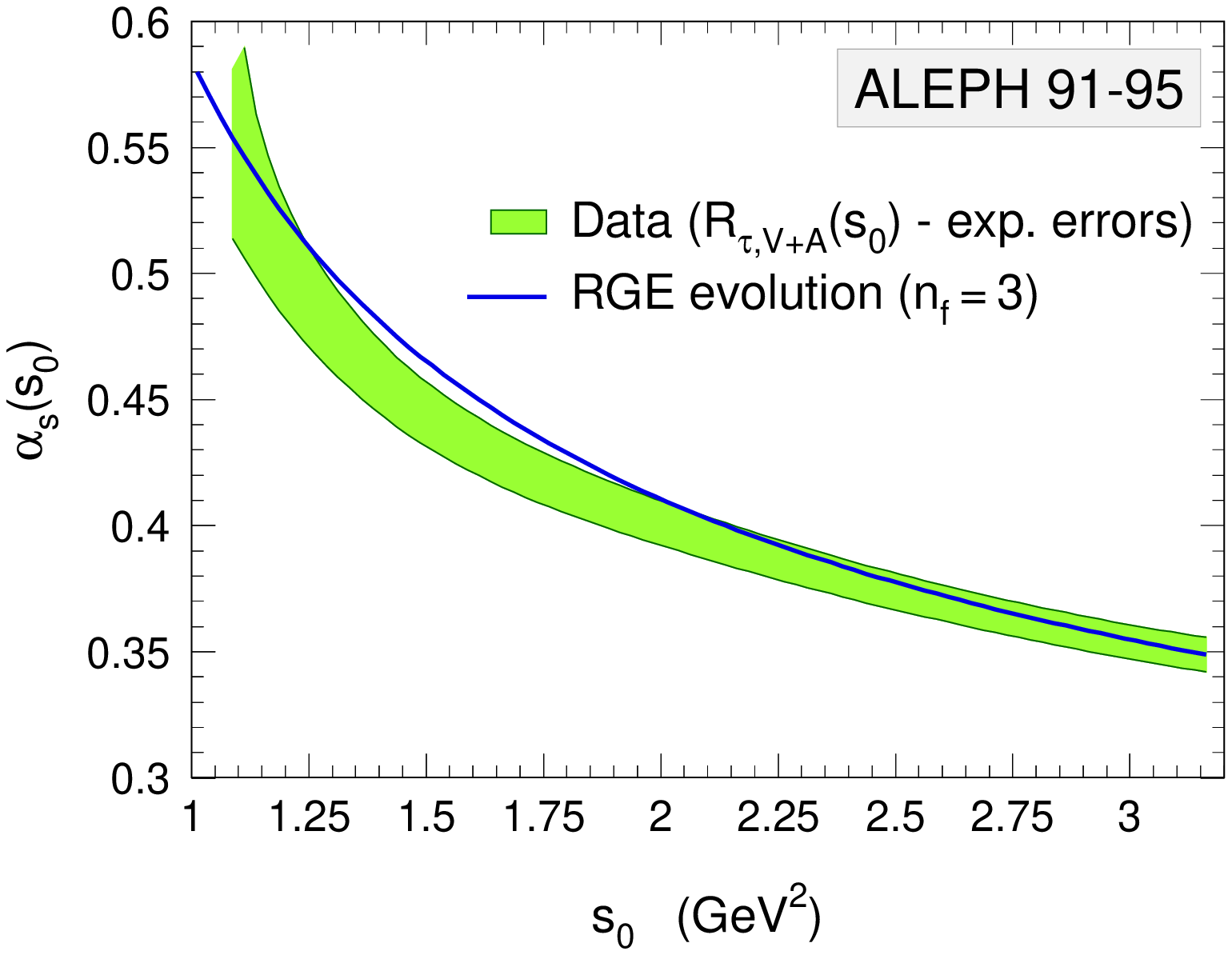}}
  \vspace{-0.8cm}
  \caption[.]{\label{vpa_runrtauas}
        \underline{Left:} 
	The ratio $R_{\tau,V+A}$ versus the square ``$\tau$ mass'' $s_0$.
      	The curves are plotted as error bands to emphasize their 
       	strong point-to-point correlations in $s_0$. Also 
      	shown is the theoretical prediction using \FOPTCI\ and
       	the results for $R_{\tau,V+A}$ and the nonperturbative 
       	terms from Table~\rm\ref{tab_asresults}.
        \underline{Right:} 
	The running of $\as(s_0)$ obtained from the 
       	fit of the theoretical prediction to $R_{\tau,V+A}(s_0)$ using
       	CIPT. The shaded band shows the data including only experimental
       	errors. The curve gives the expected four-loop RGE evolution 
       	for three flavors.}
\end{figure*}

The measured function $R_{\tau,V+A}(s_0)$ is plotted in the left
hand plot of Fig.~\ref{vpa_runrtauas} together with the theoretical 
prediction using the results of Table~\ref{tab_asresults}.  
The correlations between two adjacent points in $s_0$ are large as the 
only new information is provided by the small mass difference between the 
two points and the slightly modified weight functions under the integral. 
Moreover the correlations are reinforced by the original experimental and 
theoretical correlations. Below $1\gev^2$ the error of the theoretical 
prediction of $R_{\tau,V+A}(s_0)$ starts to blow up because of the 
increasing sensitivity to the unknown $K_4$ perturbative term;
errors of the nonperturbative contributions are {\it not} contained in 
the theoretical error band. Figure~\ref{vpa_runrtauas} (right) shows the plot 
corresponding to Fig.~\ref{vpa_runrtauas} (left), translated into the running 
of $\as(s_0)$. Only experimental errors are shown. Also plotted is the 
four-loop RGE evolution using three quark flavors.

It is remarkable that the theoretical prediction using the parameters 
determined at the $\tau$ mass and $R_{\tau,V+A}(s_0)$ extracted from 
the measured
$V+A$ \sf\ agree down to $s_0 \sim 0.8\gev^2$. The agreement is good to
about 2\% at $1\gev^2$. This result, even more directly illustrated by the
right hand plot of Fig.~\ref{vpa_runrtauas}, demonstrates the validity of 
the perturbative approach
down to masses around $1\gev$, well below the $\tau$ mass scale. The
agreement with the expected scale evolution between 1 and $1.8\gev$ is an
interesting result, considering the relatively low mass range, where
$\as$ is seen to decrease by a factor of 1.6 and reaches rather 
large values $\sim 0.55$ at the lowest masses. This behavior provides
confidence that the \asm  measurement is on solid phenomenological ground.

\subsubsection{\boldmath Final assessment on the $\as(m_\tau^2)$ determination}
\label{sec:assess}

Although this evaluation of $\as(m_\tau^2)$ represents the
state-of-the art, several remarks can be made:
\begin{itemize}

\item 	The analysis is based on the ALEPH spectral functions and branching
	fractions, ensuring a good consistency between all the observables, 
        but not exploiting the full experimental information currently 
        available from other experiments. Because the result on 
        $\as(m_\tau^2)$ 
        is limited by theoretical uncertainties, one should expect only a 
        small improvement of the final error in this way, however it can 
        influence the central value.

\item 	One example for this is the evaluation of the strange component. Some 
	discrepancy is observed between the ALEPH measurement of the 
	$(K \pi \pi)^- \nu$ mode and the CLEO and OPAL results. 
        Although this could still be the
	result of a statistical fluctuation, their average provides a 
	significant shift in the central value compared to using the ALEPH 
	number alone.
	Another improvement is the substitution of the measured branching
	fraction for the $\Km\nu$ mode by the more precise value predicted 
        from $\tau$--$\mu$ universality. Both operations have the effect 
        to increase $\RtauS$, the ratio of the $\tau$ decay width into 
        strange hadronic final states to the electronic width, from 
	$0.1603 \pm 0.0064$, as obtained by ALEPH, to $0.1686 \pm 0.0047$
	for the world average.

\item 	One can likewise substitute the world average value for the 
	universality-improved value of the electronic branching fraction,
        $\BR_e^{\rm uni}=(17.818 \pm 0.032)\%$, to the
	ALEPH result, $\BR_e=(17.810 \pm 0.039)\%$, with little change
	in the central value, but some improvement in the precision.

\end{itemize}

From this analysis, one finds the new value for the nonstrange ratio,
\beqn
\label{eq:rtau_new}
  R_{\tau,V+A} 	&=& R_\tau - R_{\tau,S} \nonumber \\
       	        &=& (3.640 \pm 0.010)-(0.1686 \pm 0.0047) \nonumber \\
                &=& 3.471 \pm 0.011~. 
\eeqn
The result~(\ref{eq:rtau_new}) translates into the following determination 
of $\as(m_\tau^2)$ from the inclusive $V+A$ component using the \FOPTCI\  
approach
\beq
\label{astau:final}
  \as(m_\tau^2) = 0.345 \pm 0.004_{\rm exp} \pm 0.009_{\rm th}~,
\eeq
with improved experimental and theoretical precision over the ALEPH
result.
Most of the theoretical uncertainty originates from the limited             
knowledge of the perturbative expansion, only predicted to third order.
Following the dicussion above we take the result from the CIPT expansion, 
not introducing any additional uncertainty spanning the difference between 
FOPT and CIPT results. The dominant theoretical errors are from the 
uncertainty on $K_4$ and from the renormalization scale dependence, both 
covering the effect of truncating the series after the estimated fourth order.

\subsubsection{\boldmath Evolution to $M_Z^2$}
\label{sec:qcd_as_evolution}

It is customary to compare \as values, obtained at different 
renormalization scales, at the scale of the $Z$-boson mass. 

The evolution of the \asm measurement from the inclusive $V+A$ 
observables given in Eq.~(\ref{astau:final}), based on Runge-Kutta 
integration of the RGE~(\ref{eq:betafun}) to N$^3$LO , and three-loop 
quark-flavor matching, gives
\beqn
\label{eq:asres_mz}
   \as(M_Z^2) &=& 0.1215~(4_{\rm exp})~(10)_{\rm th}~(5)_{\rm evol}~, 
                           \nonumber \\
                         &=& 0.1215 \pm 0.0012~.
\eeqn
The first two errors originate from the \asm determination given
in Eq.~(\ref{astau:final}). The last
error receives contributions from the uncertainties in 
the $c$-quark mass (0.00020, $m_c$ varied by $\pm0.1\gev$) 
and the $b$-quark mass (0.00005, $m_b$ varied by $\pm0.1\gev$), the 
matching scale (0.00023, $\mu$ varied between $0.7\,m_q$ 
and $3.0\,m_q$), the three-loop truncation in the matching 
expansion (0.00026) and the four-loop truncation in the RGE 
equation (0.00031), where we used for the last two errors 
the size of the highest known perturbative term as systematic 
uncertainty. These errors have been added in quadrature.
The result~(\ref{eq:asres_mz}) is a determination of the strong coupling
at the $Z$ mass scale with a precision of 1\%.

\begin{figure}[h]
%
  \centerline{\epsfxsize7.8cm\epsffile{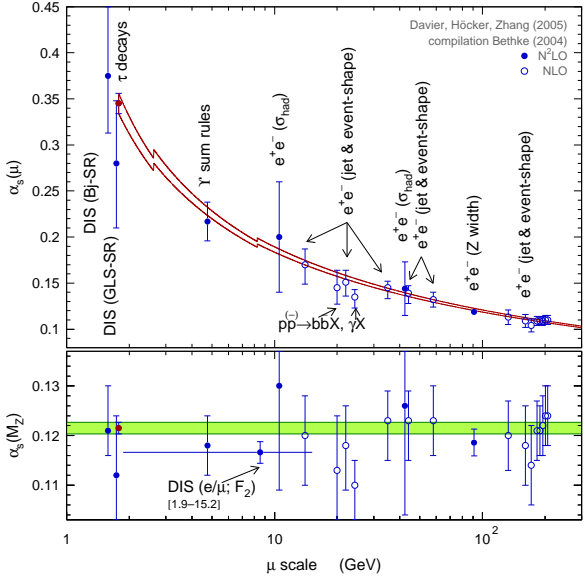}}
  \vspace{-0.6cm}
  \caption[.]{\label{fig:evolution}
      	\underline{Top}: The evolution of \asm~(\ref{astau:final}) to 
	higher scales $\mu$ using the four-loop RGE and the 3-loop 
	matching conditions applied at the heavy quark-pair thresholds 
	(hence the discontinuities at $2m_c$ and $2m_b$). The
      	evolution is compared with other independent measurements (see
      	text) covering scales varying over more than two orders magnitude. 
	The experimental values are taken from the compilation~\cite{bethke04}.
        \underline{Bottom}: The corresponding extrapolated $\as$ values
       	at $M_Z$. The shaded band displays the $\tau$ decay result within 
	errors.
	}
  \vspace{-0.4cm}
\end{figure}

The evolution path of \asm is shown in the upper plot 
of Fig.~\ref{fig:evolution}.
The two discontinuities are due to the quark-flavor matching
at $\mu=2m_q$. One could prefer to have an (almost) smooth matching 
by choosing $\mu=m_q$. However, in this case,
one must first  evolve from $m_\tau$ down to $\mbar_c$ to match the 
$c$-quark flavor, before evolving to $\mbar_b$. 
The effect on \asZ from this ambiguity is within the assigned 
systematic uncertainty for the evolution.

The comparison with the other determinations of $\as(M_Z^2)$ is given in
Fig.~\ref{fig:evolution} using compiled results from Ref.~\cite{bethke04}.

\subsubsection{\boldmath A measure of asymptotic freedom between $m_\tau^2$ and $M_Z^2$}
\label{sec:qcd_as_runningness}

The $\tau$-decay and $Z$-width determinations have comparable accuracies, 
which are however very different in nature. The $\tau$ value is 
dominated by theoretical uncertainties, whereas the determination 
at the $Z$ resonance, benefiting from the much larger energy scale 
and the correspondingly small uncertainties from the truncated 
perturbative expansion, is limited by the experimental precision 
on the electroweak observables, essentially the ratio of leptonic 
to hadronic peak cross sections. The consistency between the two results
provides the most 
powerful present test of the evolution of the strong interaction 
coupling, as it is predicted by the nonabelian nature of 
the QCD gauge theory. 
This test extends over a range of $s$ spanning more than three 
orders of magnitude. The difference between the extrapolated
$\tau$-decay value and the measurement at the $Z$ is:
\beq
 \as^\tau(M_Z^2) - \as^Z(M_Z^2) = 0.0029~(10)_\tau~(27)_Z
\eeq
which agrees with zero with a relative precision of 2.4\%.

In fact, the comparison of these two values is valuable since they are among
the most precise single measurements and they are widely spaced in energy 
scale. Thus it allows one to perform an accurate test of asymptotic freedom.
Let us consider the following evolution estimator~\cite{delphi_run} 
for the inverse of $\as(s)$,
\beq
 r(s_1,s_2) = 2\cdot\frac {\as^{-1}(s_1)-\as^{-1}(s_2)} 
                           {\ln s_1 -\ln s_2}~,
\eeq
which reduces to the logarithmic derivative of $\as^{-1}(s)$ 
when $s_1 \rightarrow s_2$, 
\beqn
\frac {d\as^{-1}}{d\ln \sqrt{s}} &=& -\frac {2\pi\beta(s)}{\as^2}~,\nonumber \\
 &=&  \frac {2\beta_0}{\pi}~\left (1 + \frac {\beta_1} {\beta_0} \frac {\as}{\pi} +\cdots~\right )~, 
\eeqn
with the notations of Eq.~(\ref{eq:betafun}). 
At first order, the logarithmic derivative is driven by $\beta_0$.

The $\tau$ and $Z$ experimental determinations of $\as(s)$ yield the value
\beq
 r_{\rm exp}(m_\tau^2,M_Z^2) = 1.405 \pm 0.053~,
\eeq
which agrees with the prediction using the RGE to N$^3$LO, 
and 3-loop quark-flavor matching,
\beq
 r_{\rm QCD}(m_\tau^2,M_Z^2) = 1.353 \pm 0.006~.
\eeq
To our knowledge this is the most precise experimental test of the
asymptotic freedom property of QCD at present. It can be compared to 
an independent  determination~\cite{delphi_run}, using an analysis of event 
shape observables at LEP between the $Z$ energy and 207~\gev,
$r(M_Z^2,(207~\gev)^2) = 1.11 \pm 0.21$, for a QCD expectation
of 1.27.

\section{Conclusions}
Using the ALEPH spectral functions and branching ratios, complemented by
other available measurements, and a revisited analysis of the 
theoretical framework, the value
$\asm = 0.345 \pm 0.004_{\rm exp} \pm 0.009_{\rm th}$ is obtained.
Taken together with the determination of \asZ from the global electroweak
fit, this result leads to the most accurate test of asymptotic freedom:
the value of the logarithmic slope of $\alpha_s^{-1}(s)$ is found to
agree with QCD at a precision of 4\%.

\end{document}